\documentclass[12pt,a4paper]{article}
\usepackage{graphics}
\usepackage{amssymb}

\begin{document}
\begin{center}{\Large \bf Coevolution of agents and networks \\ in an epidemiological
model \footnote{ This is a written version of a talk to be given at
the International School on Complexity: Course on Statistical
Physics of Social Dynamics: Opinions, Semiotic Dynamics, and
Language (Erice, 13-20 July, 2007).}}\end{center}

\begin{center}{Dami\'an H. Zanette  \\ {\it Consejo Nacional de Investigaciones Cient\'{\i}ficas y T\'ecnicas \\
Centro At\'omico Bariloche and Instituto Balseiro \\ 8400 San Carlos
de Bariloche, R\'{\i}o Negro, Argentina}}\end{center}

\begin{abstract}
We study the spreading of an infection within  an SIS
epidemiological model on a network. Susceptible agents are given the
opportunity of breaking their links with infected agents, and
reconnecting those links with the rest of the population. Thus, the
network coevolves with the population as the infection progresses.
We show that a moderate reconnection frequency is enough to
completely suppress the infection. A partial, rather weak isolation
of infected agents suffices to eliminate the endemic state.
\end{abstract}

\section{Introduction}
\label{intro}

Among the many potential applications of agent-based models whose
pattern of interactions is represented by a network, a broad class
is constituted by systems where that pattern is not a static
structure, but evolves in response to the changing state of the
agents. Generally, the evolution of the interaction network and the
dynamics of individual agents  occur over different time scales. For
instance, in  learning processes --such a those implemented in real
and artificial neural networks \cite{learning}--  connections change
adaptively over scales that are large as compared with the internal
dynamics of agents. At the opposite limit, in models of network
growth, the pattern evolves in the absence of any dynamics related
to agents at the network nodes \cite{growth}.

When, on the other hand, the dynamical time scales of a population
of agents and its interaction network are comparable, we can speak
about their {\em coevolution}. Consider, for instance, agents
involved in a cooperation-defection game, such as in the prisoner's
dilemma \cite{zim}, where defected agents are given the opportunity
of breaking the connection with their defectors, in such a way that
no further interaction is possible between them. Or, in a process of
opinion diffusion, that agents which do not succeed at reaching an
agreement may elude mutual contacts in the future
\cite{stau,gil1,gil2,holme}. In such cases, a realistic assumption
would be that the breaking of interaction links occurs just after a
few, or even one of, such events.

In this talk, we explore a model of agent-network coevolution in a
population where an infection is spreading, such that non-infected
agents are given the possibility of avoiding contact with their
infected partners --perhaps in response to risk perception
\cite{risk}. The model is based on an SIS epidemiological process,
where each agent can be susceptible (S) or infected (I). We recall
that in the standard SIS process each infected agent recovers and
becomes susceptible at a fixed rate, say, with probability $\gamma$
per unit time. Susceptible agents, in turn, become infected by
contagion from infected agents, at a rate proportional to the
infection probability per unit time, $\rho$, and to the fraction of
infected agents, $n_{\rm I}$. Within a mean-field description of the
standard SIS model, the fraction of infected agents obeys
\begin{equation} \label{sis0}
\dot n_{\rm I}  = -\gamma n_{\rm I} + \rho n_{\rm I} n_{\rm S},
\end{equation}
where $n_{\rm S}=1-n_{\rm I}$ is the fraction of susceptible agents.
In this description, for asymptotically long times, $n_{\rm I}$
vanishes if $\rho \le \gamma$. Therefore, the infection is
suppressed as time elapses. If, on the other hand, $\rho > \gamma$,
the fraction of infected agents approaches a finite value
$n_I^*=1-\gamma/\rho>0$, and the infection is endemic. The
transition between these two regimes occurs through a transcritical
bifurcation.

We introduce in the following an implementation of the SIS model on
a network, with agents occupying the nodes and contagion taking
place along the links. A susceptible agent can become infected only
if it is connected to an infected agent. Moreover, susceptible
agents have the opportunity, with a certain probability, of breaking
their connection with an infected partner before contagion takes
place, and reconnect the broken link with any other agent in the
population. We show that these reconnection events are able to
control infection spreading, even to the point of completely
suppressing the infection as the reconnection probability grows.
This suppression is achieved with a moderate decrease in the number
of network connections per infected agent.

\section{Model and mean-field formulation}
\label{model}

The present model is a variation of the model introduced by Gross
{\it et al.} \cite{thg}.
 Consider a population of $N$ agents at the nodes of a network
with $M$ links. The average number of neighbours per agent is
$z=2M/N$. At a given time, each agent can be in one of two states:
susceptible (S-agent) or infected (I-agent). Initially, the $M$
links are distributed at random over the population, and there is a
certain fraction of agents in each state.

Each evolution step is divided into two sub-steps. In the first
sub-step, an agent is chosen at random from the whole population. If
it is infected, it recovers with probability $\gamma$, and becomes
susceptible. In the second sub-step, a pair of linked agents is
chosen at random. If both agents are susceptible or infected,
nothing happens. Otherwise, the S-agent is given the chance to break
the link with its infected neighbour and reconnect it with another
agent, taken at random from the remaining of the population. This
rewiring happens with probability $r$. Finally, if the rewiring has
not occurred, the S-agent becomes infected with probability
$\lambda$.

A time unit contains $N$ evolution steps. Thus, each I-agent has a
probability $\gamma$ per time unit of becoming susceptible, so that
the mean duration of the infection period is $\gamma^{-1}$.
Moreover, each S-agent in contact with an I-agent becomes in turn
infected with probability $2 (1-r) \lambda$ per time unit.

\subsection{Mean-field equations}

Although, clearly, the model is defined bearing in mind its
implementation as a numerical simulation, a formulation in terms of
differential equations, from mean-field arguments, turns out to give
an essentially correct description of the system. To define our
mean-field formulation, we first introduce a suitable set of
variables. We call $N_{\rm I}$ and $N_{\rm S}$ the number of
infected and susceptible agents, respectively. They satisfy $N_{\rm
I}+N_{\rm S}=N$. Moreover, we  denote by $M_{\rm II}$, $M_{\rm IS}$,
and $M_{\rm SS}$, respectively, the number of network links joining
two infected agents (II), an infected agent and a susceptible agent
(IS), and two susceptible agents (SS). Since the total number of
links is preserved by the dynamical rules, we have $M_{\rm
II}+M_{\rm IS}+M_{\rm SS}=M$ constant at all times. Differential
equations will be formulated for the fractions $n_{\rm I} = N_{\rm
I}/N$, $m_{\rm II}=M_{\rm II}/M$, and $m_{\rm IS}=M_{\rm IS}/M$. The
conservation of the number of agents and links implies $n_{\rm
S}=N_{\rm S}/N = 1-n_{\rm I}$ and $m_{\rm SS}=M_{\rm SS} /M
=1-m_{\rm II}-m_{\rm IS}$.

The evolution of the above defined quantities is given by the events
of infection and recovery, and by the reconnection of links. When,
for instance, an I-agent becomes susceptible, there is not only a
decay in the fraction of I-agents, but also a change in the
fractions $m_{\rm II}$ and $m_{\rm IS}$. In fact,  the links joining
the recovered agent with I and S-agents pass, respectively, from the
II-type to the IS-type, and from the IS-type to the SS-type. A
symmetric situation occurs when an S-agent becomes infected. The
number of links of each type associated to a given agent is
calculated using mean-field-like averages. For instance, the number
of II-links associated to an I-agent is estimated as $2M_{\rm II} /
N_{\rm I} = z m_{\rm II}/n_{\rm I} $, where $z=2M/N$ is the overall
average number of links per agent. Similarly, the number of IS-links
associated to an I-agent is taken to be $M_{\rm IS} /  N_{\rm I} = z
m_{\rm IS}/2n_{\rm I}$. Reconnection events, in turn, can change an
IS-link into an SS-link. The probability for this change is
proportional to the fraction of S-agents in the whole population and
to the reconnection probability $r$. Finally, it has to be taken
into account that both infection and reconnection only occur when an
IS-link is selected, so that their probabilities are proportional to
$m_{\rm IS}$.

Putting all these considerations together, the evolution equations
for $n_{\rm I}$, $m_{\rm II}$, and $m_{\rm IS}$ read
\begin{equation} \label{sisn}
\begin{array}{ll}
n_{\rm I}' = &-n_{\rm I}+ \tilde \lambda m_{\rm IS} ,\\
m_{\rm II}' =& -2 m_{\rm II} +\tilde \lambda m_{\rm IS}^2/(1-n_{\rm I}), \\
m_{\rm IS}' =& 2m_{\rm II}-m_{\rm IS}- \tilde r (1-n_{\rm I}) m_{\rm IS} \\
& +\tilde \lambda m_{\rm IS}(2-2m_{\rm II}-3m_{\rm IS})/(1-n_{\rm
I}),
\end{array}
\end{equation}
with $\tilde \lambda = (1-r) \lambda /\gamma$ and $\tilde r =2r/ z
\gamma$. Primes indicate differentiation with respect to the
rescaled time $t'=\gamma t$. In this formulation, thus, the inverse
recovery probability $\gamma^{-1}$ fixes an overall time scale, and
the infection and reconnection probabilities are accordingly
redefined as $\tilde \lambda$ and $\tilde r$, which are the only two
parameters left. Note that $\tilde \lambda$ depends on both
$\lambda$ and $r$, and that $\tilde r$ incorporates the only
network-specific parameter, namely, the average connectivity per
agent $z=2M/N$.

In the absence of reconnection events, $r=0$, and provided that we
put $m_{\rm II}=n_{\rm I}^2$ and $m_{\rm IS}=2n_{\rm I}n_{\rm
S}=2n_{\rm I}(1-n_{\rm I})$, the three lines in Eq. (\ref{sisn})
collapse into
\begin{equation}
n_{\rm I} ' = -n_{\rm I} +2 \tilde \lambda n_{\rm I} (1-n_{\rm I}).
\end{equation}
Rewriting this equation in terms of the non-normalized parameters,
we reobtain the mean-field equation (\ref{sis0})  for the SIS model,
with $\rho=2\lambda$. The factor of $2$ relating the infection
frequencies $\rho$ and $\lambda$ originates in the fact that, in our
implementation of the SIS model on a network, each agent is (on the
average) chosen twice per unit time to possibly become infected.

\subsection{Infection level at equilibrium}

We focus the attention on the equilibrium solutions of Eqs.
(\ref{sisn}), which are the candidates to represent the infection
level and network structure at asymptotically long times. First, we
consider the stationary values of the fraction of infected agents.
The analysis is restricted to the case of $\tilde \lambda>1/2$
which, in the absence of reconnection events, corresponds an endemic
infection where a finite fraction of the population is infected at
all times, $n_{\rm I} > 0$ for $t \to \infty$.

At the fixed points of the dynamical equations (\ref{sisn}), the
equilibrium fraction of links $m_{\rm II}^*$ and $m_{\rm IS}^*$ are
related to the equilibrium fraction of infected agents $n^*_{\rm I}$
as
\begin{equation} \label{ms}
m_{\rm II}^*=\frac{n^{*2}_{\rm I}}{2\tilde \lambda (1-n^*_{\rm I})},
\ \ \ \ \ \ \ \ m_{\rm IS}^*= \frac{n^*_{\rm I}}{\tilde \lambda}.
\end{equation}
In turn, $n^*_{\rm I}$ satisfies
\begin{equation} \label{ni0}
0= n_{\rm I}^* \left[ 2\tilde \lambda -1-\tilde r +(3\tilde
r-2\tilde \lambda) n_{\rm I}^* -3\tilde r n_{\rm I}^{*2} +\tilde r
n_{\rm I}^{*3} \right].
\end{equation}
This polynomial equation has four solutions. One of them tends to
infinity for $\tilde r \to 0$, and remains real and larger than one
for any positive $\tilde r$. Since meaningful solutions to our
problem must verify $n_{\rm I}^*\le 1$, we disregard this solution
from now on.

The trivial solution $n_{\rm I}^{(0)}=0$ exists for any value of the
normalized infectivity $\tilde \lambda$ and of the normalized
reconnection probability $\tilde r$. For a given $\tilde \lambda$,
its stability depends on $\tilde r$. As discussed in more detail
below, $n_{\rm I}^{(0)}=0$ is unstable for small $\tilde r$ and
becomes stable as $\tilde r$ grows. The other two solutions read
\begin{equation} \label{n12}
n_{\rm I}^{(1,2)} = 1-\sqrt{\frac{2\tilde \lambda}{3\tilde r}}
\left[ \cos \frac{\alpha}{3} \mp \sqrt{3} \sin \frac{\alpha}{3}
\right]
\end{equation}
with
\begin{equation}
\alpha = \arctan \sqrt{\frac{32 \tilde \lambda^3}{27 \tilde r}-1}
\end{equation}
($0\le \alpha \le \pi/2$). These two solutions are real for $32
\tilde \lambda^3 \ge 27 \tilde r$. Otherwise, they are complex
conjugate numbers. The solution $n_{\rm I}^{(1)}$, with the minus
sign in the right-hand side of Eq. (\ref{n12}), approaches
$1-(2\tilde \lambda)^{-1}$ for $\tilde r \to 0$. Thus, it represents
the expected fraction of infected agents in the absence of
reconnection. When it is real, it satisfies $n_{\rm I}^{(1)}<1$, and
it is stable as long as it remains positive. Consequently, along
with the trivial solution, $n_{\rm I}^{(1)}$ is another meaningful
equilibrium solution to our problem. Finally, $n_{\rm I}^{(2)}$ is
negative and stable for small $\tilde r$. Depending on  $\tilde
\lambda$, it can become positive as $\tilde r$ grows but, at the
same time, it becomes unstable. Therefore, it does not represent a
meaningful solution.

\begin{figure}
\resizebox{\columnwidth}{!}{\includegraphics*{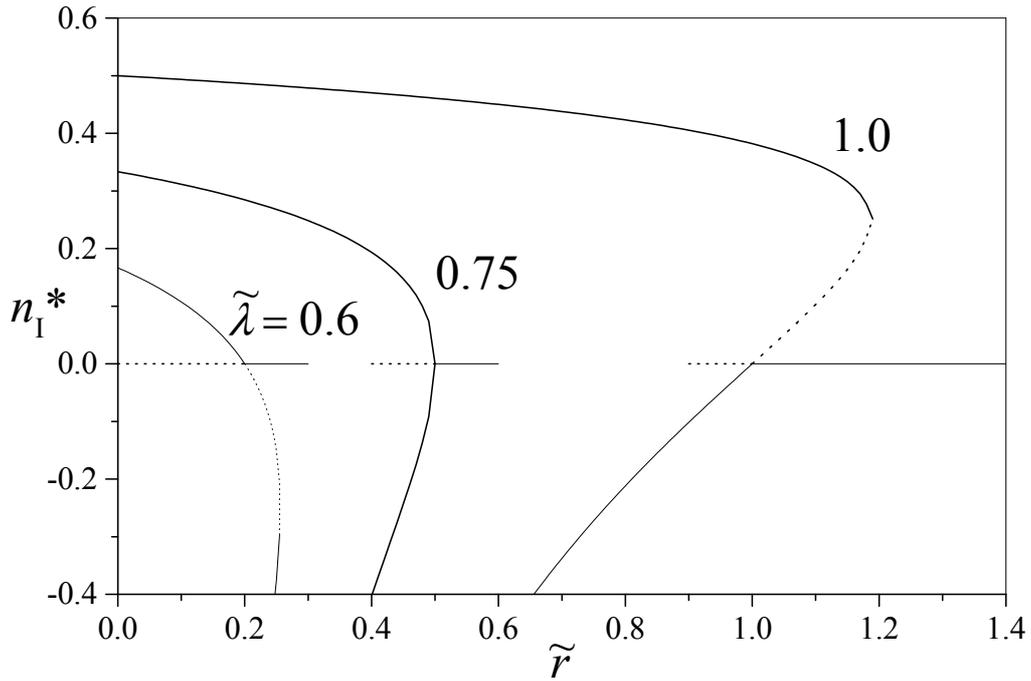}} \vspace*{15
pt}
 \caption{ Bifurcation diagram for the equilibrium fraction
of infected agents $n_{\rm I}^*$ as a function of the normalized
reconnection probability $\tilde r$, for three normalized
infectivities $\tilde \lambda$. Although only positive values of
$n_{\rm I}^*$ are meaningful, an interval in the negative domain is
also shown for completeness.  Full and dotted lines represent,
respectively, stable and unstable branches. For clarity, the
solution $n_{\rm I}^{(0)}=0$ is plotted in the vicinity of the
transcritical bifurcation only.} \label{fig1}
\end{figure}

Figure \ref{fig1} summarizes, in a bifurcation diagram, the
behaviour of $n_{\rm I}^{(0)}$, $n_{\rm I}^{(1)}$, and $n_{\rm
I}^{(2)}$ as functions of the normalized reconnection probability
$\tilde r$, for three representative values of the normalized
infectivity $\tilde \lambda$. In the three cases, we have $\tilde
\lambda>1/2$, so that --as discussed above-- a non-trivial
meaningful solution does exist. Full and dotted lines represent,
respectively, stable and unstable branches. For small infectivity
($\tilde \lambda = 0.6$), the stable solution $n_{\rm I}^{(1)}$
crosses $n_{\rm I}^{(0)}$ and becomes negative and unstable, while
$n_{\rm I}^{(0)}$ becomes stable. This transcritical bifurcation
takes place at $\tilde r = 2\tilde \lambda -1$. As $\tilde r$ grows
further,  $n_{\rm I}^{(1)}$ and the negative stable solution $n_{\rm
I}^{(2)}$ approach each other, and collide when $n_{\rm I}^{(1)} =
n_{\rm I}^{(2)}= 1-3/4\tilde \lambda$. Beyond this tangent
bifurcation, which takes place at $\tilde r = 32 \tilde \lambda^3 /
27$, the two solutions are complex numbers.

The situation is different for larger infection probabilities, as
illustrated by Fig. \ref{fig1} for $\tilde \lambda=1$. Now, for
$\tilde r=0$,  $n_{\rm I}^{(1)}$ is large and, as $\tilde r$ grows,
it is the stable negative solution $n_{\rm I}^{(2)}$ which first
reaches $n_{\rm I}^{(0)}$. At the transcritical bifurcation at
$\tilde r = 2\tilde \lambda -1$, $n_{\rm I}^{(2)}$ becomes positive
and unstable, and $n_{\rm I}^{(0)}$ becomes stable. The tangent
bifurcation where $n_{\rm I}^{(1)}$ and $n_{\rm I}^{(2)}$ collide
and become complex, at $\tilde r = 32 \tilde \lambda^3 / 27$, takes
now place when these two solutions are positive. As a consequence,
there is an interval of normalized reconnection probabilities,
between the two bifurcations, where the system is bistable: both
$n_{\rm I}^{(0)}$ and $n_{\rm I}^{(1)}$ are stable meaningful
solutions to the problem. The asymptotic state is selected by the
initial condition for $n_{\rm I}$, and $n_{\rm I}^{(2)}$ stands at
the boundary between the two attraction basins.

The regimes of small and large infectivity are separated by the
critical value $\tilde \lambda =3/4=0.75$, also shown in Fig.
\ref{fig1}. At this critical point, the transcritical and the
tangent bifurcation collapse into a  pitchfork bifurcation at
$\tilde r = 1/2$. Here, the three equilibria collide simultaneously,
and $n_{\rm I}^{(0)}$ becomes stable, while the other two solutions
become complex.

\begin{figure}
\resizebox{\columnwidth}{!}{\includegraphics*{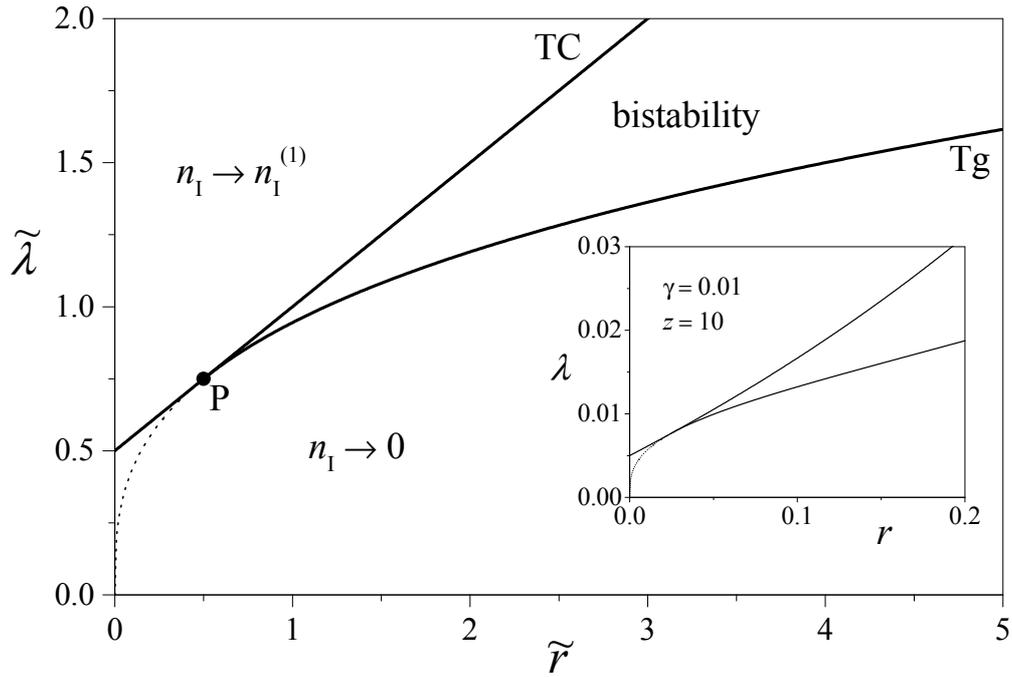}} \vspace*{15
pt} \caption{Phase diagram in the $(\tilde r, \tilde
\lambda)$-plane, showing the regions of infection suppression
($n_{\rm I} \to 0$) and persistence ($n_{\rm I} \to n_{\rm
I}^{(1)}$), and the intermediate bistability zone. Their boundaries
are given by the transcritical (TC) and the tangent (Tg) bifurcation
lines, which collapse into a pitchfork bifurcation (P) at
$(1/2,3/4)$. The dotted line is the continuation of the tangent
bifurcation line in the zone where $n_{\rm I}^{(1)}$ is negative.
The inset shows the same phase diagram in terms of the
non-normalized parameters $r$ and $\lambda$, for $\gamma=0.01$ and
$z=10$. } \label{fig2}
\end{figure}

A phase diagram of our system over the parameter plane ($\tilde
r,\tilde \lambda$) is shown in Fig. \ref{fig2}. The zones of endemic
infection, where the fraction of infected agents at asymptotically
long times is positive ($n_{\rm I}\to n_{\rm I}^{(1)}$), and of
infection suppression ($n_{\rm I}\to 0$) are separated, for large
$\tilde \lambda$ and $\tilde r$, by the bistability region, where
both asymptotic behaviours can be obtained, depending on the initial
condition. The three zones are limited by the lines of the
transcritical bifurcation [$\tilde \lambda = (1+\tilde r)/2$, TC],
where $n_{\rm I}^{(0)}=0$ changes its stability, and of the tangent
bifurcation [$\tilde \lambda = (27 \tilde r/32)^{1/3}$, Tg] where
$n_{\rm I}^{(1)}$ and $n_{\rm I}^{(2)}$ collide and become complex.
These two lines are tangent to each other at the ``triple point''
$(1/2,3/4)$, where the bistability region disappears, and the system
undergoes a pitchfork bifurcation [P]. For smaller $\tilde \lambda$
and $\tilde r$ bistability is no more possible, and the zones of
infection persistence and suppression are separated by the
transcritical line. The tangent bifurcation takes now place at
negative values of $n_{\rm I}^{(1)}$ and $n_{\rm I}^{(2)}$ (dotted
line).

The inset of Fig. \ref{fig2} shows the same phase diagram in terms
of the original, non-normalized reconnection and infection
probabilities, $r$ and $\lambda$, for a recovery probability
$\gamma=0.01$ over a network with an average of $z=10$ neighbours
per site.  Note that the relation between non-normalized and
normalized parameters is not a mere change of scale, because both
$r$ and $\lambda$ enter the definition of $\tilde \lambda$.

Let us summarize our results on the persistence or suppression of
the infection in terms of the non-normalized parameters. First, for
small infectivity, $\lambda \le \gamma/2(1-r)$, the infection is
always suppressed. In this situation, the infectivity is just too
small to sustain a finite infected population. For larger
infectivities, on the other hand, the infection can become
established, depending on the reconnection probability $r$. In the
range $\gamma/2(1-r) <\lambda <3\gamma/4(1-r)$, the infection is
endemic if reconnections are infrequent, $r<z(\lambda-\gamma/2)/(1
+z \lambda)$. Otherwise, for sufficiently frequent reconnections,
the infection dies out. The transition between both situations is
continuous in the fraction of infected agents, and occurs through a
transcritical bifurcation. For even larger infection probabilities,
$\lambda >3\gamma/4(1-r)$, the regimes of persistence (low $r$) and
suppression (large $r$) are separated by a bistability zone, where
the infection persists or dies out depending on the initial fraction
of infected agents. The bistability zone is limited by the
transcritical bifurcation quoted above and a tangent bifurcation at
a reconnection probability given by the solution to $27 \gamma^2 r=
16 (1-r)^3 \lambda^3$. The discontinuous nature of the tangent
bifurcation implies that the endemic state present in the
bistability zone disappears abruptly at the boundary, with a finite
jump in the asymptotic fraction of infected agents, from $n_{\rm I}
= 1-\sqrt{z(1-r)\lambda/3r}>0$ to zero.

\begin{figure}
\resizebox{\columnwidth}{!}{\includegraphics*{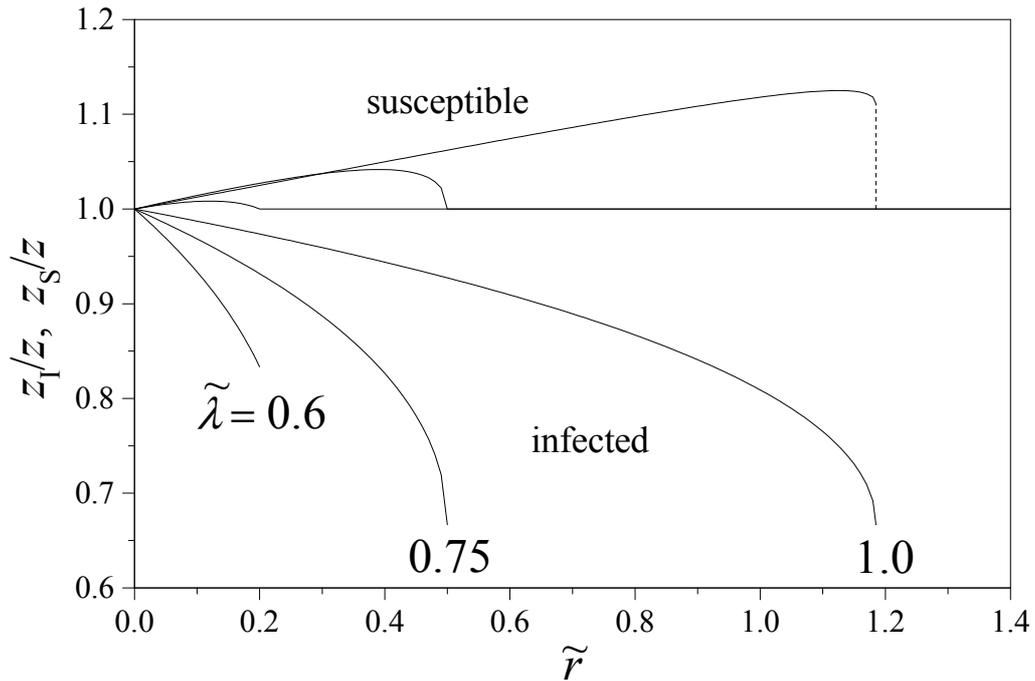}}\vspace*{15
pt} \caption{ Connectivity of infected and susceptible agents,
$z_{\rm I}$ and $z_{\rm S}$, relative to the overall connectivity
$z=2M/N$, for three values of the normalized infectivity $\tilde
\lambda$, as functions of the normalized reconnection probability
$\tilde r$.  Only the values corresponding to meaningful stable
solutions for the fraction of infected agents at plotted. The
connectivity of infected agents is not plotted beyond the threshold
of infection suppression. The vertical dashed line represents the
finite jump in $z_{\rm S}$ at the tangent bifurcation where the
solution $n_{\rm I}^{(1)}$ disappears.} \label{fig3}
\end{figure}

\subsection{Number of neighbours of infected and susceptible agents}

The variables $m_{\rm II}$ and $m_{\rm IS}$ characterize how the
structure of the network is related to the state of the agents.
Reconnection events favor the growth of the number of SS-links at
the expense of IS-links. Thus, for $r>0$, S-agents should
asymptotically posses relatively large numbers of neighbours. The
equilibrium values  $m_{\rm II}^*$ and $m_{\rm IS}^*$ as functions
of the equilibrium fraction $n_{\rm I}^*$ of I-agents are given by
Eqs. (\ref{ms}). These equations show, as expected, that the
fraction of links connecting I-agents with any other agent is
proportional to the fraction of I-agents itself.

In order to introduce quantities that define the connectivity of
I-agents and S-agents independently of their respective fractions,
we consider the average number of neighbours per agent of each type.
For I-agents, for instance, the average numbers of infected and
susceptible neighbours are $2M_{\rm II}/N_{\rm I}$ and $M_{\rm
IS}/N_{\rm I}$, respectively. The average connectivity of I-agents,
$z_{\rm I}$, is the sum of these two quantities or, equivalently,
\begin{equation}\label{zi}
\frac{z_{\rm I}}{z} = \frac{1}{2\tilde \lambda (1-n_{\rm I}^*)},
\end{equation}
which gives the ratio between $z_{\rm I}$ and the overall average
connectivity per agent at equilibrium. With analogous arguments for
S-agents, their average connectivity reads
\begin{equation} \label{zs}
\frac{z_{\rm S}}{z} = \frac{1}{1-n_{\rm I}^*}-\frac{n_{\rm
I}^*}{2\tilde \lambda (1-n_{\rm I}^*)^2} .
\end{equation}
Due to the conservation of the total number of links, $z_{\rm I}$
and $z_{\rm S}$ are univocally related. This relation can be
obtained from Eqs. (\ref{zi}) and (\ref{zs}) by eliminating $n_I^*$,
which yields
\begin{equation}
z_{\rm S} = z_{\rm I}[1+2\tilde \lambda (1-z_{\rm I}/z)].
\end{equation}
In order to describe the correlation between the structure and the
state of the population it is however useful to analyze both $z_{\rm
I}$ and $z_{\rm S}$ as functions of the relevant parameters. Figure
\ref{fig3} illustrates the behaviour of $z_{\rm I}$ and $z_{\rm S}$,
as described in the following, for the infection probabilities
$\tilde \lambda$ already considered in Fig. \ref{fig1}.

For $n_{\rm I}^* = n_{\rm I}^{(1)}$, which stands for the stable
equilibrium solution for low reconnection probabilities, both
$z_{\rm I}$ and $z_{\rm S}$ approach $z$ as $\tilde r \to 0$. As
expected, in the absence of reconnection events, there is no
difference in the number of neighbours of infected and susceptible
agents. Also, as $\tilde r$ grows from zero, we have $z_{\rm
I}<z<z_{\rm S}$. We verify that reconnection tends to increase the
connectivity of S-agents at the expense of I-agents.

The other solution relevant to the epidemiological process, $n_{\rm
I}^*=n_{\rm I}^{(0)}=0$, corresponds to a purely susceptible
population. Accordingly, we find $z_{\rm S}=z$. Note also that Eq.
(\ref{zi}) predicts $z_{\rm I} = z/2\tilde \lambda$, but this value
is never realized due to the total absence of I-agents in this
state.

For $\tilde \lambda \le 3/4$, the fraction of I-agents decreases
monotonically with $\tilde r$ and vanishes continuously at the
transcritical bifurcation --or, for $\tilde \lambda = 3/4$, at the
pitchfork bifurcation. The connectivity of S-agents is $z_{\rm S}
=z$ both at $\tilde r=0$ and at the bifurcation. For intermediate
values of the reconnection probability $z_{\rm S}$ is larger than
$z$ and attains a maximum. This maximum, which at first sight may
result to be surprising, can be easily explained. In fact, to
sustain a value of $z_{\rm S}$ larger than the overall average $z$,
it is necessary to have I-agents with a relatively low number of
neighbours. As the infection is progressively suppressed by
reconnection, the number of I-agents decreases and, accordingly,
their contribution to the average number of neighbours per agent
becomes less significant. At the bifurcation and beyond, S-agents
must account for the whole average, so that $z_{\rm S}$  returns to
its value for $\tilde r=0$, i.e. $z_{\rm S}=z$.

The connectivity of I-agents, in turn, is a monotonically decreasing
function of $\tilde r$, and reaches $z_{\rm I}=z/2\tilde \lambda <z$
at the bifurcation. This implies that, even at the threshold of
infection suppression, I-agents maintain a finite number of
neighbours within the population.

For $\tilde \lambda > 3/4$, again, the connectivity $z_{\rm S}$
associated with the solution $n_{\rm I}^{(1)}$ initially increases
with $\tilde r$, and attains a maximum. In the subsequent decay,
however, it does not reach $z_{\rm S}=z$. In fact, $n_{\rm I}^{(1)}$
disappears through a tangent bifurcation when it is still positive,
so that the jump in the infection level is discontinuous. At the
bifurcation, we find  $z_{\rm S}= 2z(2\tilde \lambda +3)/9>z$. The
connectivity of I-agents decreases with $\tilde r$ and, at the
bifurcation, its value is independent of $\tilde \lambda$: $z_{\rm
I}=2/3$.

From the viewpoint of the interplay of the epidemiological dynamics
and the structure of the underlying network, the most interesting
result of this section is the fact that the infection dies out even
when infected agents keep a substantial connectivity with the rest
of the population. In the cases illustrated in Fig. \ref{fig3}, for
instance, infected agents preserve more than $60$ \% of their
connections at the threshold where the infection level vanishes. In
other words, reconnection needs not to completely isolate infected
agents to suppress the infection. A moderate, partial isolation of
the infected population is enough to asymptotically inhibit the
endemic state.

\section{Discussion and conclusion}

What mechanisms are at work when the infection is suppressed by
reconnection, even when the connectivity of infected agents remains
fairly high? To advance an answer to this question, it helps to
consider a simpler dynamical system for the fraction of I-agents:
\begin{equation} \label{eff}
n_{\rm I}'=-n_{\rm I} +[2 \tilde \lambda - \tilde r (1-n_{\rm I})^2
]n_{\rm I} (1-n_{\rm I}).
\end{equation}
The right-hand side of this equation is just a rearrangement of that
of Eq. (\ref{ni0}). The equilibria of Eq. (\ref{eff}) are thus
identical to the equilibria for $n_{\rm I}$ in Eqs. (\ref{sisn}).
Moreover, their stability properties are also the same as in our
original system. It is important to understand, however, that
(\ref{eff}) and (\ref{sisn}) are not equivalent: they merely share
the same equilibrium behaviour in which regards the fraction of
I-agents.

We immediately see that Eq. (\ref{eff}) can be put in the form of
the standard mean-field equation (\ref{sis0}) for a SIS process if
we introduce the effective infection probability
\begin{equation}
\rho_{\rm eff} = \gamma [2 \tilde \lambda - \tilde r (1-n_{\rm I})^2
]= 2 (1-r)\lambda -\frac{2r}{z} (1-n_{\rm I})^2.
\end{equation}
In Eq. (\ref{sis0}) the threshold of infection suppression, where
the trivial equilibrium  changes its stability, is given by
$\rho=\gamma$. Imposing this same condition to $\rho_{\rm eff}$, we
find $\tilde r = 2\tilde \lambda-1$. But this is precisely the
suppression threshold in the system with reconnections. Therefore,
with respect to the stabilization of the trivial equilibrium, the
system (\ref{sisn}) is effectively equivalent to the standard SIS
model with infectivity $\rho_{\rm eff}$. The transcritical
bifurcation of Eqs. (\ref{sisn}), where $n_{\rm I}^{(0)}=0$ becomes
stable, can be interpreted as a kind of continuation for $r \neq 0$
of the transcritical bifurcation of the SIS model without
reconnection events.

\begin{figure}
\resizebox{\columnwidth}{!}{\includegraphics*{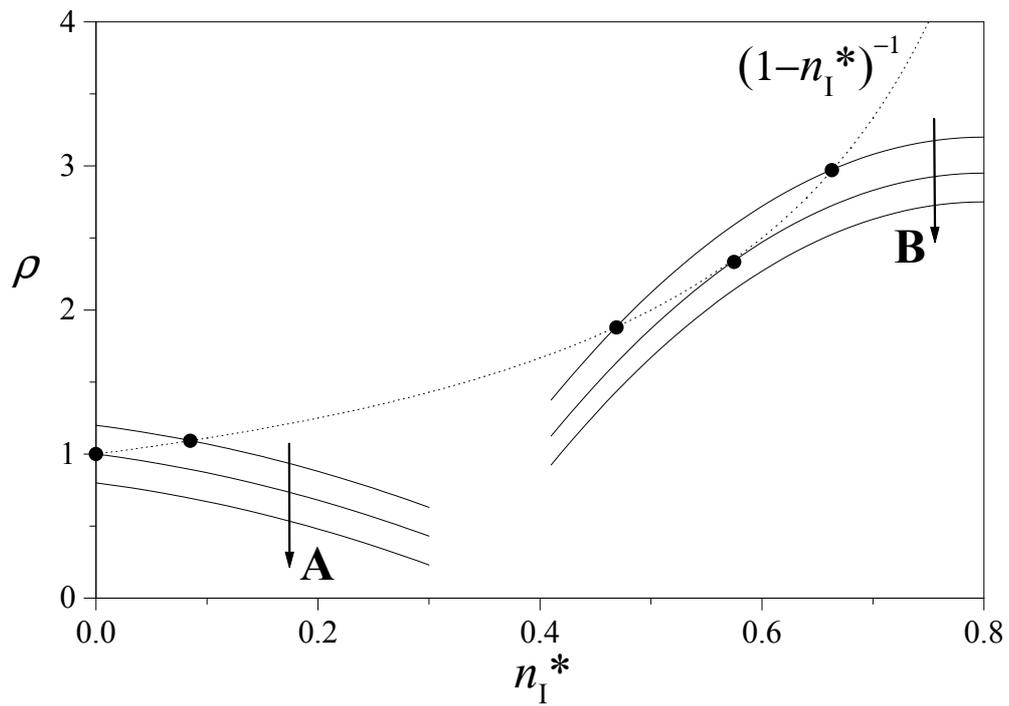}}\vspace*{15
pt} \caption{Graphical solution of Eq. (\ref{eff1}). The dotted
curve represents the right-hand side of the equation, and full
curves are possible graphs of the left-hand side. Dots stand at
their intersections. The arrows illustrate how the graphs may change
upon the variation of parameters, in the cases of a transcritical
bifurcation (A) and of a tangent bifurcation (B).} \label{fig4}
\end{figure}

The interpretation of the tangent bifurcation where the endemic
state disappears at a positive value of $n_{\rm I}^{(1)}$, for
$\tilde \lambda >3/4$, is less direct. It can however be argued that
the presence of such a tangent bifurcation, together with the
transcritical bifurcation which stabilizes the trivial equilibrium,
constitute the generic critical behaviour expected for a SIS model
like Eq. (\ref{eff}), with an infection probability which depends on
the density of I-agents:
\begin{equation} \label{eff0}
n_{\rm I}'=-n_{\rm I} + \rho (n_{\rm I}) n_{\rm I} (1-n_{\rm I}).
\end{equation}
Besides the trivial equilibrium, this equation has fixed points at
the solutions of
\begin{equation} \label{eff1}
\rho (n_{\rm I}^*) = (1-n_{\rm I}^*)^{-1}.
\end{equation}
Figure \ref{fig4} illustrates graphically two representative
situations. The dotted curve is the graph of the right-hand side of
Eq. (\ref{eff1}) as a function of $n_{\rm I}^*$. If the graph of
$\rho (n_{\rm I}^*)$ has a single intersection with the dotted curve
(A) and if, upon variation of parameters in the infection
probability, the graph varies as indicated by the arrow, the
intersection crosses $n_{\rm I}^*=0$ and a transcritical bifurcation
takes place. The standard SIS model, for which $\rho$ is constant,
is a special case within this situation. More generally, the graph
of $\rho (n_{\rm I}^*)$ may have two (B) or more intersections with
the dotted curve. When parameters change, it is still possible than
one of the intersections becomes involved in a transcritical
bifurcation crossing $n_{\rm I}^*=0$, exactly as in situation A.
Now, however, it may well be the case that two intersections
approach each other, and eventually collapse and disappear, as in B.
In this case, Eq. (\ref{eff0}) undergoes a tangent bifurcation, as
found to happen in our system (\ref{sisn}).

In summary, the above discussion shows that the suppression of the
endemic state as a result of reconnection events can be reasonably
understood in terms of the critical behaviour of a standard SIS
model with an effective infectivity, which depends on both the
reconnection probability and on the fraction of infected agents. The
transcritical bifurcation which stabilizes the state where the
infection is completely inhibited is interpreted as a continuation
of a similar transition in the absence of reconnection. In turn, the
tangent bifurcation --which, for large infectivities, suppresses the
infection as the reconnection probability grows-- is a generic
phenomenon in SIS models with density-dependent infectivity
\cite{tangent}.

\centerline{ } \centerline{***} \centerline{ }

We have studied a model for an epidemiological process in a
population of agents on a network, where contagion can occur along
the network links. The network coevolves with the population as the
infection progresses: a susceptible agent can decide to break a link
with an infected partner, and reconnect it with another agent. Our
main result is that this reconnection mechanism, implemented with
moderate frequency, can completely suppress the endemic state where
an infected portion of the population persists at arbitrary long
times. Suppression of the endemic state does not require full
isolation of the infected population. On the contrary, it can be
achieved while each infected agent preserves a substantial part of
the links with the rest of the population.

Whether these results are relevant to real epidemiological processes
is a question beyond the present study. Our analysis should be
regarded as an illustration of the kind of collective effects that
may emerge from the coevolution of populations of dynamical elements
and their interaction network.

\end{document}